\documentclass{article}

\usepackage{epstopdf}
\usepackage{epsfig}
\usepackage{booktabs}

\usepackage{microtype}
\usepackage{graphicx}
\usepackage{subfigure}
\usepackage{booktabs} 
\usepackage{enumitem} 
\usepackage{standalone}
\usepackage{pgfplots}
\usepackage{pgfplotstable}
\usepackage{pifont}
\usepackage{float}
\usepackage{multirow}
\usepackage{hyperref}
\usepackage{verbatim}
\usepackage{wrapfig}
\usepackage{appendix}
\usepackage{rotating}
\usepackage{mathtools}
\usepackage{xcolor}
\usepackage{bbm}
\usepackage{tikz}
\usepackage{tabu}
\usepackage{xintexpr}
\usepackage{siunitx}
\usepackage[flushleft]{threeparttable}
\usetikzlibrary{tikzmark,calc}

\usepackage{amsmath,amsfonts,bm, amsthm}

\def\Figref#1{Fig.~\ref{#1}}

\def\eqref#1{equation~\ref{#1}}
\def\Eqref#1{Equation~\ref{#1}}

\def\1{\bm{1}}

\def\rvm{{\mathbf{m}}}

\def\rvw{{\mathbf{w}}}
\def\rvx{{\mathbf{x}}}
\def\rvy{{\mathbf{y}}}

\def\rmC{{\mathbf{C}}}

\def\rmW{{\mathbf{W}}}

\DeclareMathAlphabet{\mathsfit}{\encodingdefault}{\sfdefault}{m}{sl}
\SetMathAlphabet{\mathsfit}{bold}{\encodingdefault}{\sfdefault}{bx}{n}

\newcommand{\R}{\mathbb{R}}

\theoremstyle{plain}

\theoremstyle{remark}

\theoremstyle{definition}

\theoremstyle{plain}

\theoremstyle{plain}

\theoremstyle{definition}
\newtheorem{definition}{Definition}
\newtheorem{theorem}{Theorem}
\newtheorem{corollary}{Corollary}

\newtheorem{lemma}{Lemma}

\providecommand{\corollaryname}{Corollary}
\providecommand{\lemmaname}{Lemma}
\providecommand{\problemname}{Problem}
\providecommand{\remarkname}{Remark}
\providecommand{\theoremname}{Theorem}

\usepackage{hyperref}

\usepackage[accepted]{icml2021}

\icmltitlerunning{Selective Network Linearization}

\begin{document}

\twocolumn[
\icmltitle{Selective Network Linearization for Efficient Private Inference}



\icmlsetsymbol{equal}{*}

\begin{icmlauthorlist}
\icmlauthor{Minsu Cho}{to}
\icmlauthor{Ameya Joshi}{to}
\icmlauthor{Siddharth Garg}{to}
\icmlauthor{Brandon Reagen}{to}
\icmlauthor{Chinmay Hegde}{to}
\end{icmlauthorlist}

\icmlaffiliation{to}{New York University Tandon School of Engineering, New York}

\icmlcorrespondingauthor{Minsu Cho}{mc8065@nyu.edu}

\icmlkeywords{Machine Learning, ICML}

\vskip 0.3in
]



\printAffiliationsAndNotice{}  

\begin{abstract}

Private inference (PI) enables inference directly on cryptographically secure data.
While promising to address many privacy issues, it has seen limited use due to extreme runtimes.
Unlike plaintext inference, where latency is dominated by FLOPs, in PI non-linear functions (namely ReLU) are the bottleneck. Thus, practical PI demands novel ReLU-aware optimizations.
To reduce PI latency we propose a gradient-based algorithm that selectively linearizes ReLUs while maintaining prediction accuracy. 
We evaluate our algorithm on several standard PI benchmarks. The results demonstrate up to $4.25\%$ more accuracy (iso-ReLU count at 50K) or $2.2\times$ less latency (iso-accuracy at 70\%) than the current state of the art and advance the Pareto frontier across the latency-accuracy space.
To complement empirical results, we present a ``no free lunch" theorem that sheds light on how and when network linearization is possible while maintaining prediction accuracy. Public code is available at \url{https://github.com/NYU-DICE-Lab/selective_network_linearization}.


\end{abstract}


\section{Introduction}

Cloud-based machine learning frameworks motivate the setting of private inference (PI). At a high level, the vision of private inference is to enable a user to (efficiently) perform inference of their data on a model owned by a cloud service provider. But both parties wish to preserve their privacy and both the user's data and the service provider's model are encrypted prior to inference using cryptographic techniques. 

Curiously, the difficulty in realizing this vision is the nonlinear operations of a deep network. Private execution of the linear operations in a network using ciphertext can be made essentially as fast as normal (plaintext) evaluation of the same operations with secret sharing techniques and input independent preprocessing phases. However, to privately evaluate ReLUs in the network, some version of Yao's Garbled Circuits (GC) is necessary, resulting in proportionally high latency costs and storage overheads. In a nutshell: standard deep network architectures (such as ResNets) are ill-suited for efficient PI, since \emph{they contain far too many ReLUs}. See~\cite{DELPHI,ghodsi2020,ghodsi2021circa,DEEPREDUCE} and a detailed discussion below in Section~\ref{sec:background}. 

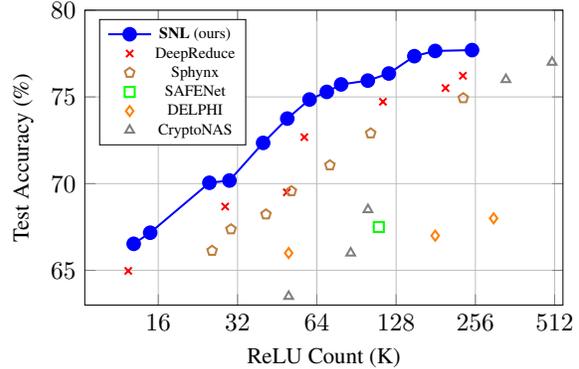
\begin{figure}[!t]
    \centering
    \resizebox{0.95\linewidth}{!}{
    \begin{tikzpicture}
    \begin{semilogxaxis}[legend pos=north west,
        legend style={nodes={scale=0.62, transform shape}},
        width=\columnwidth,
        height=5.7cm,
        xlabel= \small ReLU Count (K),
        ylabel=\small Test Accuracy (\%),
        xlabel style={at={(0.5, 0.1)}},
        ylabel style={at={(-0.05, 0.5)}},
        xmin=0, xmax=550,
        ymin=63,ymax=80,
        ylabel near ticks,
        xlabel near ticks,
        axis background/.style={fill=blue!0},
        grid=both,
        log basis x = 2,
        /pgf/number format/1000 sep={\,},
        log ticks with fixed point,
        grid style={line width=.1pt, draw=gray!10},
        major grid style={line width=.2pt,draw=gray!50},
        ]
    \addplot[mark=*,mark size=2.5pt,blue, thick] 
        plot coordinates {
            (12.9, 66.53)
            (14.9, 67.17) 
            (25.0, 70.05)
            (29.8, 70.18)
            (40.0, 72.35)
            (49.4, 73.75)
            (60.0, 74.85)
            (69.8, 75.29)
            (79.1, 75.72)
            (99.9, 75.94)  
            (120.0, 76.35)
            (150.0, 77.35)
            (180.0, 77.65)
            (248.4, 77.7)
        };
    \addlegendentry{\textbf{SNL} (ours)}


    

    \addplot[mark=x,mark size=2pt,red, thick, only marks] 
        plot coordinates {
            (12.3, 64.97)
            (28.7 , 68.68)
            (49.2, 69.50)
            (57.34, 72.68)
            (114.0, 74.72)
            (197.0, 75.51)
            (229.38,76.22)
        };
    \addlegendentry{DeepReduce}


    \addplot[mark=pentagon, mark size=2pt, brown, thick, only marks] 
        plot coordinates {
            (25.6, 66.13)
            (30.2, 67.37)
            (41.0, 68.23)
            (51.2, 69.57)
            (71.7, 71.06)
            (102.4, 72.90)
            (230, 74.93)
        };
    \addlegendentry{Sphynx}

    \addplot[mark=square,mark size=2pt,green, thick, only marks] 
        plot coordinates {
            (110, 67.5)
        };
    \addlegendentry{SAFENet}

    \addplot[mark=diamond,mark size=2pt,orange, thick, only marks] 
        plot coordinates {
            (50, 66)
            (180, 67)
            (300, 68)
        };
    \addlegendentry{DELPHI}
    
    \addplot[mark=triangle,mark size=2pt, gray, thick, only marks] 
        plot coordinates {
            (50, 63.5)
            (86, 66)
            (100, 68.5)
            (334, 76)
            (500, 77)
        };
    \addlegendentry{CryptoNAS}
    \end{semilogxaxis}
\end{tikzpicture}}
    \caption{Our approach, SNL, achieves the Pareto frontier on ReLU counts vs.\ test accuracy for CIFAR-100.}
    \label{fig:C100-pareto-curve}
\end{figure}

Thus, unlocking the full potential of fast, accurate, and private neural inference require rethinking network design to have as few ReLUs as possible.
Numerous efforts in this direction have already emerged. Prior work such as MiniONN~\cite{liu2017oblivious} focus on the security protocols themselves, while more recent works such as Delphi~\cite{DELPHI} or Circa~\cite{ghodsi2021circa} propose to replace ReLUs with other activations. An alternative line of work designs ReLU-efficient network skeletons using neural architecture search (NAS). CryptoNAS~\cite{ghodsi2020} uses evolutionary NAS, while Sphynx~\cite{cho2021sphynx} uses micro-search NAS.

Our work in this paper pursues a new path, and applies to several architectures that are used in imaging, computer vision, or other perception problems (specifically, deep convolutional networks with/without residual connections and ReLU activations). For these networks we study a problem that we call \emph{deep network linearization}.  This refers to the process of judiciously picking a subset of neurons in such a network and {eliminating} their nonlinearities (i.e., replacing their ReLU operation with an identity, or linear, operation) minimizing tradeoff between overall performance and number of ReLU operations.
Note that via this process, we are \emph{not} reducing the number of parameters in the network. Since ReLU activations are simple scalar operations, in normal scenarios we are essentially {not getting any reduction in FLOP counts} during network training or testing. However, in the PI setting, a major advantage of linearization is due to reducing the number of GC computations.

This paper introduces a simple gradient-based algorithm (that we call \emph{Selective Network Linearization}, or \textsc{SNL}) to solve the deep network linearization problem, validates this algorithm on a variety of standard benchmarks, and provides theoretical evidence that illuminates its observed behavior.




\begin{table*}[!th]
    \centering
    \begin{tabular}{c  c  c  c}
        Approach & Methods & Reduce ReLUs & Units that are removed \\ 
        \toprule 
        \multicolumn{1}{l|}{CryptoNAS~\cite{ghodsi2020}} & \multicolumn{1}{c|}{NAS} & \multicolumn{1}{c|}{Yes} & layers \\
        \multicolumn{1}{l|}{Sphynx~\cite{cho2021sphynx}} & \multicolumn{1}{c|}{NAS} & \multicolumn{1}{c|}{Yes} & layers \\
        \hline
        \multicolumn{1}{l|}{DELPHI~\cite{DELPHI}} & \multicolumn{1}{c|}{NAS + polynomial approx.} & \multicolumn{1}{c|}{Yes} & layers \\
        \multicolumn{1}{l|}{SAFENet~\cite{SAFENET}} & \multicolumn{1}{c|}{NAS + polynomial approx.} & \multicolumn{1}{c|}{Yes} & channels  \\
        \hline
        \multicolumn{1}{l|}{Unstructured Pruning} & \multicolumn{1}{c|}{N/A} & \multicolumn{1}{c|}{No} & not exist \\
        \multicolumn{1}{l|}{Structured Pruning} & \multicolumn{1}{c|}{N/A} & \multicolumn{1}{c|}{Yes} & channels, layers \\
        \hline 
        \multicolumn{1}{l|}{DeepReDuce\cite{DEEPREDUCE}} & \multicolumn{1}{c|}{manual} & \multicolumn{1}{c|}{Yes} & layers \\
        \multicolumn{1}{l|}{SNL (ours)} & \multicolumn{1}{c|}{gradient-based} & \multicolumn{1}{c|}{Yes} & pixels, channels
    \end{tabular}
    \caption{Comparison of various techniques that reduce ReLU operations in deep networks. NAS stands for neural architecture search. ``Pruning'' techniques eliminate entire neurons. \textsc{SNL}, our proposed gradient-based network linearization method, achieves the accuracy-latency Pareto frontier in private inference. 
    }
    \label{table: summary}
\end{table*}

Currently, the state-of-the-art in private inference is achieved by DeepReDuce~\cite{DEEPREDUCE}. The authors introduce a concept known as \emph{ReLU dropping} in deep networks that, at a conceptual level, is equivalent to network linearization. Their proposed algorithm consists of three stages: ReLU ``culling", ReLU ``thinning", and ReLU ``reshaping", each of which requires several manual design choices, involving several hyper-parameters. The search space of candidate architectures is very large; applying DeepReDuce to a ResNet with $D$ stages would require training $\Omega(D)$ different networks and picking the one with the best accuracy. Also, crucially, the keep/drop decisions made by DeepReDuce are made in \emph{stages}: either entire ReLU layers are linearized, or entire ReLU layers are retained as is.

In contrast, our proposed technique, \textsc{SNL}, is {highly-automated, involving very few hyper-parameters}. SNL is implemented using a single gradient-based training procedure (which we describe below in detail) and requires {no searching over multiple network skeletons}. Finally, SNL provides {fine-grained control}, all the way down to the pixel feature map level, of which ReLUs to retain or eliminate.  

The intuition underlying \textsc{SNL} is simple and applicable to many deep networks. Consider any standard architecture (e.g., a ResNet-34), except with a twist. All ReLUs are now replaced with \emph{parametric ReLUs} (or pReLUs)~\cite{he2015delving} with an \emph{independent, trainable slope parameter} for \emph{each ReLU} in the network. Further, each slope parameter is further constrained to be binary (0 or 1). Having defined this network, $\textsc{SNL}$ proceeds to perform standard training (using, say, SGD or Adam); there is no other manual intervention necessary.

Some care is needed to make things work. The main challenges lie in (a) enforcing the binary constraints on the slopes of the pReLUs, and (b) ensuring that only a small number of ReLUs are retained, i.e., the vector of slope parameters is (anti) sparse. \textcolor{black}{Mirroring the approach adopted by certain network pruning algorithms~\cite{Lee2019SNIPSN,cho2021espn}, we resolve these difficulties by augmenting the standard train error loss with an $\ell_1$-penalty term defined over the slope coefficients, decaying the weight of this penalty on a schedule, and applying a final rounding step that binarizes the slopes.} See Section~\ref{sec: problem formualtion}.

We validate \textsc{SNL} on a variety of benchmark datasets commonly used in the literature on private image classification, and show that \emph{\textsc{SNL} achieves Pareto dominance over the entire accuracy-latency tradeoff curve over all existing approaches}. Figure~\ref{fig:C100-pareto-curve} above demonstrates this in the context of CIFAR-100; we provide several additional results (and ablation studies) below and also show the same benefits for CIFAR-10 and Tiny ImageNet. 
See Section~\ref{sec:exp} and Appendix.

Probing into the results of \textsc{SNL} reveals curious behavior. It appears that ReLUs in earlier layers are selectively linearized at a (much) higher rate than ReLUs in later layers; this aligns with the conclusions arrived at by earlier PI works such as DeepReDuce \cite{DEEPREDUCE}. Could this be indicative of some fundamental property of network learning? 
In a step towards rigorously answering this question, we prove \emph{no free lunch} theorems showing that linearization comes at the cost of reducing memorization capacity of 3-layer networks. Moreover, if the network is \emph{contractive}, i.e., if the second layer has fewer neurons than the first layer (typical in classification-type scenarios), then selective linearization retains original network capacity \emph{only if} fewer neurons in the second layer are linearized. See Section~\ref{subsec: expressivity}.


\section{Related Work}
\label{sec:background}



\noindent\textbf{Private inference.}
Prior work on private inference (PI) have proposed methods that leverage existing cryptographic primitives for evaluating the output of deep networks. Cryptographic protocols can be categorized by choice of ciphertext computation used for linear and non-linear operations in a network. Operations are computed using some combination of: (1) secret-sharing (SS) under a secure multi-party computation (MPC)~\cite{shamir1979share, goldreich2019play} model; (2) partial homomorphic encryptions (PHE)~\cite{gentry2011implementing}, which allow limited ciphertext operations (e.g. additions and multiplications), and (3) garbled circuits (GC)~\cite{yao1982secure, yao1986generate} that rely on specialized circuitry. 

In this paper, our focus is exclusively on the DELPHI protocol for private inference\footnote{We chose DELPHI as a matter of convenience; the general trends discovered in our work hold regardless of the encryption protocol. We acknowledge that choosing the ``best'' protocol is an important (and fast advancing) area of research, and that better PI protocols, such as~\cite{cryptflow2}, may exist.}. 
DELPHI assumes the threat model that both parties are honest-but-curious. Therefore, each party strictly follows the protocol, but may try to learn information about the other party's input based on the transcripts they receive from the protocol. This threat model is standard in prior PI works, including MiniONN~\cite{liu2017oblivious}, DELPHI, CryptoNAS~\cite{ghodsi2020}, SAFENet~\cite{SAFENET}, and DeepReDuce~\cite{DEEPREDUCE}.

DELPHI is a hybrid protocol that combines cryptographic primitives such as secret sharing (SS) and homomorphic encryptions (HE) for all linear operations, and garbled circuits (GC) for ReLU operations. DELPHI divides the inference into two phases to make the private inference happen: the offline phase and an online phase. DELPHI's cryptographic protocol allows for front-loading all input-independent computations to an offline phase. By doing so, this enables ciphertext linear computations to be as fast as plaintext linear computations while performing the actual inference.

For non-linear (especially ReLU) operations, all the above approaches leverage GC's for secure computation. However, such circuits cause major latency bottlenecks. \citet{DELPHI} show empirical evidence that ReLU computation requires $90\%$ of the overall private inference time for typical deep networks. As a remedy, DELPHI and SAFENET~\cite{SAFENET} propose neural architecture search (NAS) to selectively replace ReLUs with polynomial operations. CryptoNAS~\cite{ghodsi2020},  Sphynx~\cite{cho2021sphynx} and DeepReDuce~\cite{DEEPREDUCE} design new ReLU efficient architectures by using macro-search NAS, micro-search NAS and multi-step optimization respectively. 


\noindent\textbf{Network compression.} 
We emphasize that network linearization is \emph{distinct} from neural network compression, which fundamentally focuses on reducing the number of \emph{learnable parameters} in a deep network.
Broadly, network compression (or pruning) techniques can be divided into two buckets. \emph{Unstructured} pruning approaches identify and remove ``unimportant'' weights (edges) in a deep network to reduce model size. \citet{zhu2017prune, gale2019state} retain top-$k$ parameters by measuring the importance of weight parameters based on their absolute magnitude. \citet{Lee2019SNIPSN} proposes measuring importance according to gradient magnitudes. \citet{xiao2019autoprune, cho2021espn} trains the network with iterative multi-objective loss functions and reparameterized weights.


In contrast, \emph{structured} pruning approaches remove entire channels in convolutional architectures; this has the effect of eliminating both nodes and edges in the network. \citet{li2016pruning} uses a combination of magnitude-pruning and fine-tuning to prune channels in every layer. 
\citet{he2020learnedFP} achieves state-of-the-art pruning performance in this direction. While structured pruning methods do end up reducing ReLU counts as a by-product, this reduction comes at the cost of losing expressivity. On the other hand, selective linearization  retains the original number of learnable parameters while only reducing the number of ReLUs in the network. Below we show extensive experiments (and preliminary theory) showing why selective linearization is superior in the context of private inference.

\color{black}


\begin{figure*}[!t]
    \centering
    \setlength{\tabcolsep}{15pt}
    \resizebox{\textwidth}{!}{
    \begin{tabular}{c c}
         \includegraphics[width=0.70\textwidth]{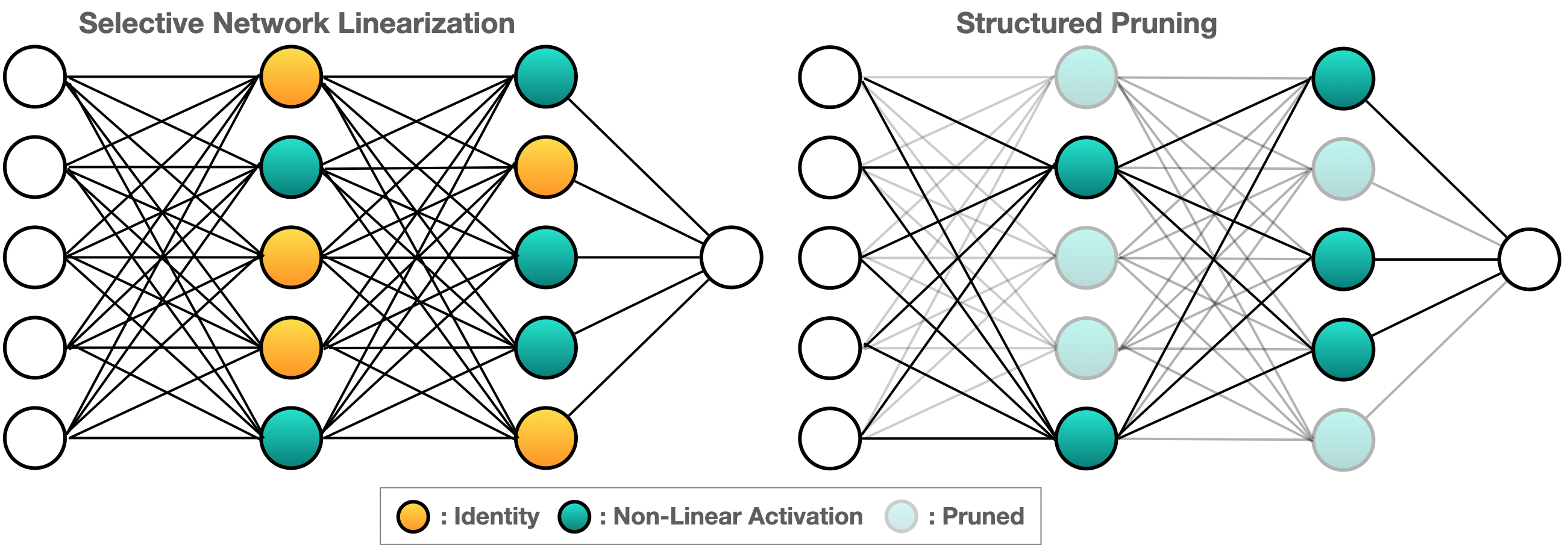} &
         \includegraphics[width=0.15\textwidth]{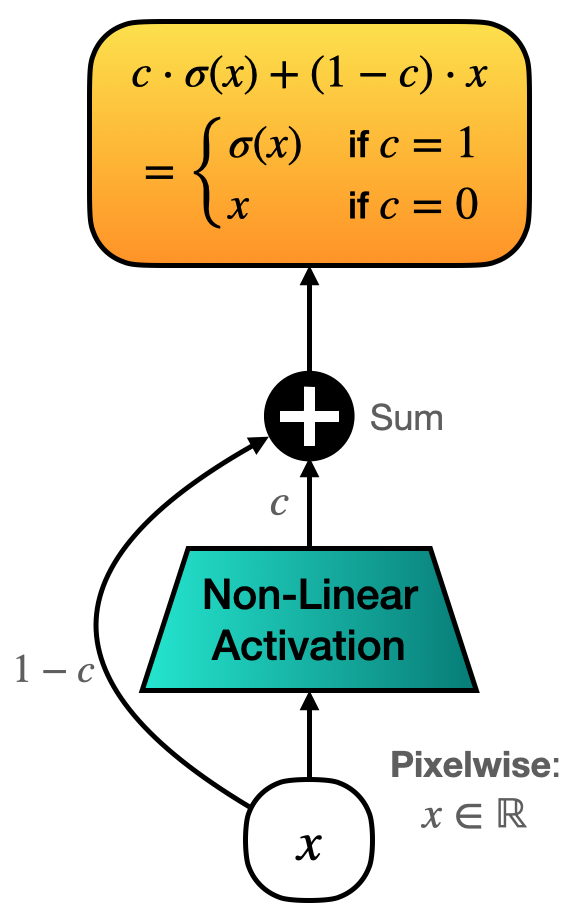}\\
         (a) Visual comparison on SNL and Structured Pruning & (b) Visual representation of Equation~\ref{eq: relu linearization}
    \end{tabular}}
    \caption{Visualization of SNL, structured pruning, and Equation~\ref{eq: relu linearization}. (a) Both SNL and structured pruning have two non-linear activations. While all 55 parameters are on in SNL, the network from structured pruning has only 18 parameters. We note that number of non-linear activations (especially ReLU) is what matters in PI. (b) Visual representation of the convex combination between $x$ and $\sigma(x)$. If non-linear activation $\sigma$ is ReLU and $c \in \R$, then this convex combination is equivalent to PReLU.}
    \label{fig: SNL vs SP and ReLU linearization}
\end{figure*}

\section{Selective Network Linearization} 
\label{sec: problem formualtion}


\paragraph{Notation.} 
We represent matrices and tensors with upper-case variables unless stated otherwise. Elements of vectors and matrices are represented using appropriate subscripts. $\|\cdot\|_0$ represents number of non-zeros; $\odot$ represents element-wise products. $[d]$ is the natural number set $\{1,2,3,\ldots, d\}$. $\mathbf{1}$ is the all-ones vector of the appropriate dimension.

\subsection{Setup}
\color{black}
Our approach applies to any feed-forward deep network architecture with $d$ layers and ReLU activations ($\sigma(\cdot)$). Let $f_{\rmW}$ be any such network where $\rmW = \{W^0,W^1,\ldots,W^d\}$ are the layer weight. Denote the pre-activation outputs of each linear layer by $z^1, z^2, \ldots, z^d$. Denote $x$ be the input vector and $z^1 = W^1 x$. Let $a^i = \sigma(z^i)$ be the outputs after each layer of ReLUs. Expanding $f_{\rmW}(x)$, we get
\begin{equation}
    f_{\rmW}(x) = W^d(\sigma(W^{d-1}(\sigma(W^{d-2}(\cdots\sigma(W^1x)\cdots).
\label{eq: deep net}
\end{equation}
We now introduce a set of auxiliary binary vectors $\rmC = \{c^1, c^2, \ldots c^d\}$, where for ${i\in [d]}$, we define $\sigma_{c^i}$ as the \emph{parametric ReLU} (PReLU) activation function (with (left) slope equal to $1-c^i$) defined for \emph{every neuron} in the network. Then, the output, $\sigma_{c^i}(\cdot)$, at the $i^\textsuperscript{th}$ layer can be formally expressed as:
\begin{equation}
    a_{c^i}^i = \sigma_{c^i}(z^i) = c^i \odot \sigma(z^i) + (\mathbf{1}-c^i) \odot z^i
    \label{eq: relu linearization}
\end{equation}
Notice that according to our definition, $a_{c^i}^i = \text{ReLU}(z^i)$ for $c_i=1$, and $a_{c^i}^i = z^i$ for $c_i = 0$. 
In essence, we are re-defining $\sigma$ to $\sigma_{c^i}$ for all $i \in [d]$ in \eqref{eq: deep net}, so that now the  feed-forward deep network $f_{\rmW, \rmC}$ is parameterized both by $\rmW$ and $\rmC$. 

Our goal is to learn a model with high accuracy but with as few ReLUs as possible. This motivates the following $\ell_0$-constrained optimization problem:
\begin{align}
    \min_{\rmW, \rmC} L(f_{\rmW, \rmC}(\rvx), \rvy) \quad \text{s.t.}~\sum_{i=1}^d\|c_i\|_0 \leq B,~c_i~\text{binary}.
    \label{eq: ReLU optimization}
\end{align}
In terms of optimization difficulty, the main challenges with \Eqref{eq: ReLU optimization} are in enforcing the binary constraints on each entry of $c_i$, and the overall $\ell_0$ constraint. Below, we propose a solution to these challenges.
\color{black}

\subsection{The SNL Algorithm}
We first drop the binary constraints, and relax the optimization problem in \Eqref{eq: ReLU optimization} to a Lasso-style objective function~\cite{tibshirani1996regression}:
\begin{equation}
    \min_{\rmW, \rmC} L(f_{\rmW, \rmC}(\rvx), \rvy) + \lambda \left(\sum_{i=1}^d \|c_i\|_1\right).
    \label{eq: relaxed relu optimization}
\end{equation}
In principle, both the weights $\rmW$ and the slope parameters $\rmC$ can be updated via standard gradient iterations over this objective. In practice, to ensure good performance, our approach requires four steps: (1) Start with a pre-trained network $f_{\rmW}$. (2) Update $\rmW$ and $\rmC$ simultaneously by gradient descent over \Eqref{eq: relaxed relu optimization}. (3) Binarize elements of $\rmC$ by rounding to 0 or 1. (4) freeze $\rmC$ and perform final finetuning of $f_{\rmW}$.

\begin{algorithm}[t!]
    \small
    \caption{\textsc{SNL}: Selective Network Linearization}
    \begin{algorithmic}[1]
        \STATE\textbf{Inputs: } $f_{\rmW}$: pre-trained network, $\lambda$: Lasso coefficient, $\kappa$: scheduling factor, $B$: ReLU budget, $\epsilon$: threshold. 
        \STATE \textcolor{black}{Set $\rmC=1$: same dimensions to all feature maps.} 
        \STATE $\overline{\rmW} \gets (\rmW, \rmC)$
        \WHILE {$\text{ReLU Count} > B$}
            \STATE Update $\overline{\rmW}$ via ADAM for one epoch.
            \STATE ReLU Count $\leftarrow \|\mathbbm{1}(\rmC > \epsilon)\|_0$
            \IF {ReLU count not decreased}
                \STATE Increment Lasso coefficient $\lambda \leftarrow \kappa \cdot \lambda$.
            \ENDIF
        \ENDWHILE
        \STATE $\rmC \leftarrow \mathbbm{1}(\rmC > \epsilon)$
        \STATE Freeze $\rmC$ and finetune $f_{\rmW}$.
    \end{algorithmic}
    \label{alg:finetune}
\end{algorithm}
 
In Step 2, we first initialize $\rmC$ to all ones in the vector space. Then, we update $\rmW$ and $\rmC$ simultaneously via standard backpropagation (using ADAM), until the desired sparsity level of $\rmC$ is achieved. We monitor the sparsity of $\rmC$ by computing the quantity $\|\mathbbm{1}(\rmC > \epsilon)\|_0$, where $\epsilon$ is a very small non-negative hyperparameter. Similar to the technique of \emph{homotopy continuation} in Lasso, if the sparsity of $\rmC$ increases compared to previous epochs we increase $\lambda$ with a small multiplicative factor $\kappa$ and repeat. Once the desired number of nonzero elements in $\rmC$ reached, we binarize by thresholding $\mathbbm{1}(\rmC > \epsilon)$. Finally, we freeze $\rmC$ and fine-tune the network weights $\rmW$ to boost final performance. Algorithm~\ref{alg:finetune} provides detailed pseudocode. 

\subsection{Aside: Linearization versus pruning} 

Let us pause to highlight the essential diffences of \Eqref{eq: ReLU optimization} with unstructured network pruning~\cite{zhu2017prune, Lee2019SNIPSN, cho2021espn}. There, one typically chooses a \emph{masking} parameter for every \emph{weight} in the network, and solves a similar looking optimization:
\begin{equation}
    \min_{\rvw, \rvm} L(f(\rvw \odot \rvm;\rvx), \rvy) \quad \text{s.t. } \|\rvm\|_0 \leq k
    \label{eq: objective function for pruning}
\end{equation}
where $k$ is a parameter counting the total number of retained weights.

But eliminating (individual) edges do not lead to reduction in ReLU counts; only if \emph{all} incoming weights to a given ReLU operation are removed -- or if an entire channel is removed -- then one can safely eliminate the corresponding ReLU operation in the feature map. This is akin to \emph{structured} pruning, but as we will show below, this leads to significantly worse performance as well as representation capacity compared to \textsc{SNL}. Figure~\ref{fig: SNL vs SP and ReLU linearization}(a) visualizes the difference between SNL and structured pruning.

\color{black}

\begin{figure*}[!t]
    \centering
    \resizebox{\textwidth}{!}{
\usetikzlibrary{pgfplots.groupplots}

\begin{tikzpicture}
    \begin{groupplot}[group style={group size=3 by 1}, cycle list name=color list,
 	xlabel={ReLU Count (in K)},
 	legend cell align=center,
 	legend columns=9, 
 	legend style={at={(0.5,1.2)},anchor=north},
 	grid=major,
 	ylabel near ticks,
 	major grid style={dashed},
 	every group plot/.append style={ultra thick},
 	every group plot/.append style={mark size={0pt}}, width=7.5cm]
 	
    \nextgroupplot[ylabel={Test Accuracy}, xmode=log, log basis x=2, 
    /pgf/number format/1000 sep={\,}, log ticks with fixed point,
    title=CIFAR-10, title style={at={(0.5, 0.95)}},]
    
    \addplot+[mark=*,mark size=2pt,blue, thick] 
        plot coordinates {
            (12.9, 88.23)
            (14.9, 88.43) 
            (25.0, 90.88)
            (29.8, 90.92)
            (40.0, 91.68)
            (49.4, 92.27)
            (60.0, 92.63)
            (69.8, 93.02)
            (79.1, 93.16)
            (99.9, 93.50)
            (150.0, 94.26)
            (180.0, 94.78)    
            (300, 95.06)
            (400, 95.07)
            (500, 95.21)
            
        };
    
    \addplot+[mark=x,mark size=2pt,red, thick, only marks] 
        plot coordinates {
            (36.0 , 88.5)
            (70.0, 90.0)
            (80.0, 90.5)
            (114.0, 92.7)
            (147.0, 93.16)
            (221.48,94.07)
        };
    
    
    \addplot+[mark=triangle,mark size=2pt, gray, thick, only marks] 
        plot coordinates {
            (50, 90.0)
            (86, 91.5)
            (100, 92.2)
            (334, 94)
            (500, 94.8)
        };
    
    \addplot[mark=oplus, mark size=2pt, cyan!80!black, thick, only marks]
        plot coordinates{
            (152.21, 84.36)
            (110.59, 78.73)
            (211.584, 84.04)
            (301.44, 85.68)
            (337.92, 86.4)
            (442.36, 85.83)
        };
        
    \addplot[mark=otimes, mark size=2pt, violet, thick, only marks]
        plot coordinates{
            (311.7, 93.34)
        };
    
    \nextgroupplot[xmode=log, log basis x=2, 
    /pgf/number format/1000 sep={\,}, log ticks with fixed point,
    title=CIFAR-100, title style={at={(0.5, 0.95)}},]
    
    \addplot+[mark=*,mark size=2pt,blue, thick] 
        plot coordinates {
            (12.9, 66.53)
            (14.9, 67.17) 
            (25.0, 70.05)
            (29.8, 70.18)
            (40.0, 72.35)
            (49.4, 73.75)
            (60.0, 74.85)
            (69.8, 75.29)
            (79.1, 75.72)
            (99.9, 75.94)  
            (120.0, 76.35)
            (150.0, 77.35)
            (180.0, 77.65)
            (248.4, 77.7)
        };
    \addlegendentry{SNL}
    
    \addplot[mark=x,mark size=2pt,red, thick, only marks] 
        plot coordinates {
            (12.3, 64.97)
            (28.7 , 68.68)
            (49.2, 69.50)
            (57.34, 72.68)
            (114.0, 74.72)
            (197.0, 75.51)
            (229.38,76.22)
        };
    \addlegendentry{DeepReduce}
    
    \addplot[mark=pentagon, mark size=2pt, brown, thick, only marks] 
        plot coordinates {
            (25.6, 66.13)
            (30.2, 67.37)
            (41.0, 68.23)
            (51.2, 69.57)
            (71.7, 71.06)
            (102.4, 72.90)
            (230, 74.93)
        };
    \addlegendentry{Sphynx}
    
    
    \addplot[mark=triangle,mark size=2pt, gray, thick, only marks] 
        plot coordinates {
            (50, 63.5)
            (86, 66)
            (100, 68.5)
            (334, 76)
            (500, 77)
        };
    \addlegendentry{CryptoNAS}
    
    \addplot[mark=square,mark size=2pt,green, thick, only marks] 
        plot coordinates {
            (110, 67.5)
        };
    \addlegendentry{SAFENet}
    
    \addplot[mark=+,mark size=2pt,yellow!80!black, thick, only marks] 
        plot coordinates {
            (12.9, 56.83)
            (14.9, 56.92)
            (25.0, 60.88)
            (29.8, 62.20)
            (40.0, 61.98)
            (49.4, 63.69)
            (60.0, 64.89)
            (69.8, 64.98)
            (79.1, 65.50)
            (99.9, 66.50)
        };
    \addlegendentry{SNIP}

    \addplot[mark=diamond,mark size=2pt,orange, thick, only marks] 
        plot coordinates {
            (50, 66)
            (180, 67)
            (300, 68)
        };
    \addlegendentry{DELPHI}
    
    \addplot[mark=oplus, mark size=2pt, cyan!80!black, thick, only marks]
        plot coordinates{
            (147.9, 52.09)
            (211.84, 54.74)
            (426.638, 60.12)
        };
    \addlegendentry{$\ell_1$ Filter Pruning}
    
    \addplot[mark=otimes, mark size=2pt, violet, thick, only marks]
        plot coordinates{
            (311.7, 70.83)
        };
    \addlegendentry{LFPC}
	
	
     

	
	
     
     
    \nextgroupplot[xmode=log, log basis x=2, 
    /pgf/number format/1000 sep={\,}, log ticks with fixed point,
    title=Tiny-ImageNet, title style={at={(0.5, 0.95)}},]
    
    \addplot[mark=*,mark size=2pt,blue, thick] 
        plot coordinates {
            (60, 54.24)
            (100, 58.94)
            (200, 63.39)
            (300, 64.04)
            (400, 63.83)
            (500, 64.42)
        };

    \addplot[mark=x,mark size=2pt,red, thick, only marks] 
        plot coordinates {
            (57.35, 53.75)
            (98.3, 55.67)
            (114.69, 56.18)
            (200.0, 57.51)
            (230.0, 59.18)
            (400.0, 61.65)
            (459.0, 62.26)
            (917.0, 64.66)
        };
        
    \addplot[mark=pentagon, mark size=2pt, brown, thick, only marks] 
        plot coordinates {
            (102.4, 48.44)
            (204.8, 53.51)
            (286.7, 56.72)
            (491.5, 59.12)
            (614.4, 60.76)
        };
	
	
     
     
    \end{groupplot}
\end{tikzpicture}	}
    \caption{SNL achieves Pareto frontiers of ReLU counts versus test accuracy on CIFAR-10, CIFAR-100, and Tiny-ImageNet. SNL outperforms the state-of-the-art methods (DeepReDuce, SAFENet, and CryptoNAS) in all range of ReLU counts on all three dataset. $\ell_1$ Filter Pruning~\cite{li2016pruning} and LFPC~\cite{he2020learning} are structured pruning techniques.}
    \label{fig: pareto curve}
\end{figure*}
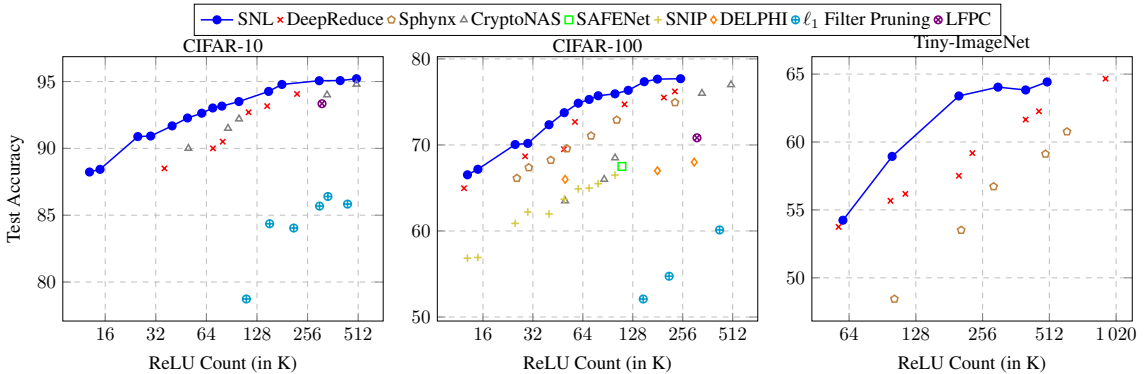

\section{Experiments}
\label{sec:exp}

\paragraph{Architectures, datasets, training.} We apply SNL to the ResNet18/34 architectures\cite{he2016deep} and Wide-ResNet 22-8 \cite{zagoruyko2016wide} architectures. As standard in prior work~\cite{DEEPREDUCE}, we remove the ReLU layers in the first convolution layer. This layer serves to raise the input channel dimension from 3 to a higher number (e.g. 64 for ResNet18/34). 


We focus on standard image classification datasets (CIFAR-10/100 and Tiny ImageNet). The image resolutions on both CIFAR-10 and CIFAR-100 is $32 \times 32$; Tiny-ImageNet is $64 \times 64$. CIFAR10 has $10$ output classes with $5000$ training images and $1000$ test images per class, while CIFAR-100 has 100 output classes with $500$ training images and $100$ test images per class. TinyImageNet has 200 output classes with 500 training images and 50 validation images. . 

We first pre-train networks on CIFAR-10/100 using SGD with initial learning rate 0.1 and momentum 0.9, decay the learning rate at 80 and 120 epochs with \color{black} 0.1 learning rate decay factor, 0.0005 weight decay, \color{black} and use batch-size 256. For Tiny-ImageNet, we use the same hyperparameters, except that we train for 100 epochs and perform learning rate decay at 50 and 75 epochs with decay factor 0.1. 

For the SNL algorithm, we initialize $\lambda=0.00001$ and increment $\lambda$ by a multiplicative factor $1.1$ if the ReLU count increases compared to previous epochs. We set the thresholding parameter $\epsilon = 0.01$. We use the ADAM optimizer with learning rate equual to $0.001$. 
In the finetuning stage, we update the network parameter $W$ via SGD with a learning rate of $0.001$ and 0.9 momentum for 100 epochs. Finally, like DeepReDuce, we use knowledge distillation during finetuning~\cite{hinton2015distilling} with temperature parameter $4$ and equal relative weights on both cross-entropy (on the hard labels) and KL divergence (on the soft labels). We use the original pretrained model with identical network topology as the teacher. 

\paragraph{Latency estimates for private inference.} We report results both in terms of ReLU counts and wall-clock time for private inference. Existing open-source implementations of DELPHI are unfortunately not compatible with networks with pixel-wise parametric ReLU operations. Therefore, we propose a method to {empirically estimate} the online latency for private inference over such networks per input data sample. We break down the sources of latency into two categories: ciphertext linear operations and ciphertext ReLU operations. Since ciphertext linear operations in DELPHI are essentially as fast as plaintext linear operations (and since ReLUs in plaintext are essentially free), we simply report the latency of ciphertext linear operations as the same as that for one plaintext network inference. For ciphertext ReLU operations, we experimentally measure the wall-clock time for $1000$ ReLU operations using DELPHI, resulting in $t = 0.021$ seconds per $1000$ ReLUs \footnote{We verify our estimates with independent prior work. For example, Table 4 of DeepReDuce~\cite{DEEPREDUCE} lists number of ReLUs and latency for several models. By performing a linear regression between these covariates and measuring the slope, we can get an estimate of the online latency per ReLU as $t = 0.019$ seconds, which is almost exactly what we get.}. 

\paragraph{Pareto analysis and comparisons.} Our main result, Figure~\ref{fig: pareto curve}, shows that SNL achieves the ReLU count-accuracy Pareto frontier on CIFAR-10, CIFAR-100, as well as Tiny-ImageNet over all previous competing approaches. 

To our knowledge, the previous best approach is due to DeepReDuce, which (as we discussed in the Introduction) optimizes networks in three stages, each with a fair bit of manual intervention. On the other hand, \textsc{SNL} outperforms DeepReDuce with a fairly straightforward gradient-based procedure. Furthermore, we used identical hyperparameters for all target ReLU budgets and base networks, showing that $\textsc{SNL}$ is capable of giving robust state-of-the-art performance without careful hyperpararameter tuning.

We now discuss our results in detail in two regimes of interest: high-ReLU and low-ReLU budgets (Table~\ref{tab:cifar100}). At a high-ReLU budget, SNL achieves 76.35\% test accuracy with budget=120K on CIFAR-100, while DeepReDuce achieves 75.50\% accuracy given a significantly higher (197K) ReLU count. Therefore, the network produced by SNL cuts down latency to $60\%$ of that of DeepReDuce, while achieving $0.85\%$ higher accuracy. In comparison, the network from CryptoNAS with a comparable performance as \textsc{SNL} requires nearly \emph{3 times} the total ReLU counts. Similarly, the network produced by Sphynx~\cite{cho2021sphynx} with  achieves $74.93\%$ with ReLU count=230K; \textsc{SNL} matches this performance with approximately half the ReLU count. On Tiny-ImageNet, \textsc{SNL} produces a network that only uses $55\%$ of ReLUs as that of DeepReduce (500K vs. 917K) with par accuracy (64.42\% vs. 64.66\%).

At low-ReLU budgets, the SNL network achieves 73.75\% test accuracy with 49.4K ReLU count on CIFAR-100. Both DeepReDuce and Sphynx use around 50K ReLU count, and achieve 69.50\% and 69.57\% test accuracy respectively. CryptoNAS network with 100K ReLU count reaches 68.75\% test accuracy, but SNL beats CryptoNAS by nearly 5\% in terms of accuracy with only half its ReLU budget. 
At the extreme case, we see that the \textsc{SNL} network with ReLU count 25K achieves $70.05\%$, outperforming DeepReDuce models with 68.68\% containing ReLU count 28.6K. On Tiny-ImageNet, SNL achieves 63.39\% with ReLU count 200K, whereas DeepReDuce network reaches 61.65\% accuracy with nearly double the ReLU count (400K). Furthermore, Sphynx achieves 4\% less accuracy with $3\times$ the ReLU counts (614K), compared to \textsc{SNL} with ReLU count 200K.

\begin{figure*}[!t]
    \centering
    \resizebox{0.9\textwidth}{!}{
    \begin{tabular}{c c c} 
         \includegraphics[width=0.25\textwidth]{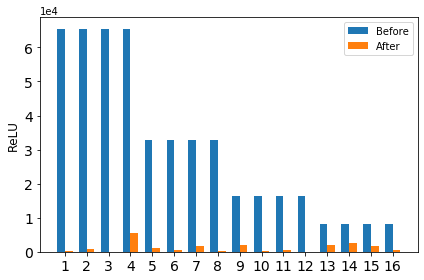} &
         \includegraphics[width=0.25\textwidth]{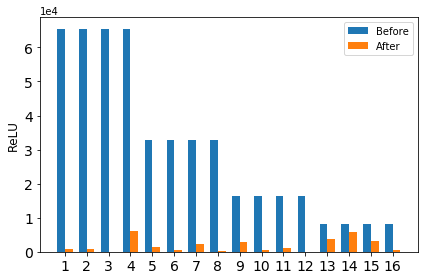} & 
         \includegraphics[width=0.25\textwidth]{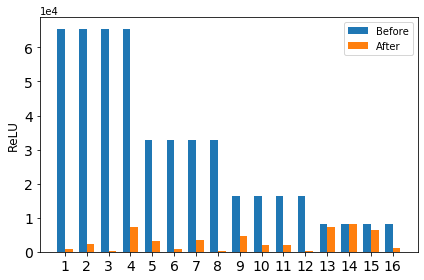} \\
         \scriptsize ReLU 20K & \scriptsize ReLU 30K & \scriptsize ReLU 50K \\
         \includegraphics[width=0.25\textwidth]{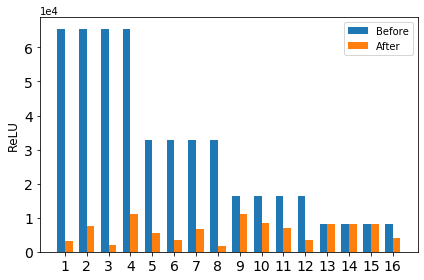} &
         \includegraphics[width=0.25\textwidth]{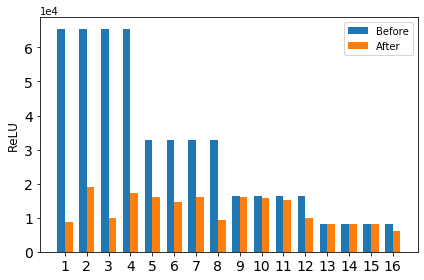} & 
         \includegraphics[width=0.25\textwidth]{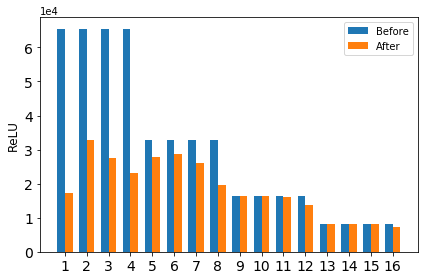} \\
         \scriptsize ReLU 100K & \scriptsize ReLU 200K & \scriptsize ReLU 300K \\
    \end{tabular}}
    \caption{ReLU distribution on ResNet18 from SNL. \textit{Before} and \textit{After} stand for ReLU counts before SNL and after SNL, respectively. $x$-axis correpsonds to a layer index of the network. The larger gap between a blue and orange bar means the more ReLU operations replaced with identity function. Aligning to the theoretical results from Corollary~\ref{corollary: optimal relu distribution}, SNL replaces more ReLU operations to identity function in the earlier layers (1-4) while later layers' ReLU operations preserved (13-16).}
    \label{fig: C100 relu distribution}
\end{figure*}

\vspace{-1em}

\paragraph{Analysis of networks produced by \textsc{SNL}.} In \textsc{SNL} we only impose a total ReLU budget constraint; the algorithm automatically decides how it wants to allocate its budget across different layers. To study the effects of this allocation, we run \textsc{SNL} on ResNet-18 trained on CIFAR-100 with various ReLU budgets ranging from 20K to 300K; the base ResNet-18 requires approximately 492K ReLU operations. Comparing the blue and the orange bars in Figure~\ref{fig: C100 relu distribution}, we see that SNL drops earlier ReLUs at a higher rate than later ReLUs; this indicates that ReLU operations on later layers are generally more crucial than the earlier layers. Indeed, the ReLU distribution plots with ReLU budget 100K, 200K, and 300K show that the ReLU operations in later layers (13, 14, 15, and 16) are mostly preserved. In contrast, the ReLUs in the earlier layers (1, 2, 3, 4) are largely eliminated and replaced with the identity operation.  This observation mirrors the approach in DeepReDuce. There, the authors discovered by exhaustive manual ablation studies that ReLUs in later layers had a more significant effect than earlier layers, and therefore proposed a ``ReLU criticality metric'' to decide whether or not to cull entire layers of ReLUs. Interestingly, \textsc{SNL} arrives at a similar strategy without any manual intervention\footnote{It appears that the authors of DeepReDuce measured criticality in ResNet \emph{stages} and not individual layers. They found that the order of ResNet18 stages on CIFAR-100 as $S_1 < S_4 < S_2 < S_3$, where $S_i$ means a group of residual blocks (e.g., $S_3$ includes 9, 10, 11, and 12\textsuperscript{th} layers in ResNet-18)}. We revisit this in Section~\ref{subsec: expressivity}.

\paragraph{Channel-wise \textsc{SNL}.} As described, the \textsc{SNL} algorithm proposes linearizing ReLUs at the finest (pixel-level) scale. We can develop ``channel-wise'' extensions of this method with a straightforward change-of-variables. Consider a CNN architecture, where our decision to linearize is at the channel (filter) scale. In this case, everything is the same as before, except that $\rmC$ in Algorithm~\ref{alg:finetune} is the same dimension as the total number of convolution filters. Let $x_i^d$ be the feature map after applying the $i\textsuperscript{th}$ filter in the $d$\textsuperscript{th} layers. We can selectively linearize this via: 
\begin{equation}
    z_i^d = c_i^d \cdot ReLU(x_i^d) + (1 - c_i^d) \cdot (x_i^d)
\end{equation}
and the rest of the algorithm proceeds identically. We call this \emph{Structured SNL} and test this with ResNet-18 and Wide-ResNet 22-8 on the CIFAR-100 dataset. Table~\ref{tab: structured SNL} in the Appendix shows that structured SNL outperforms SAFENet with 0.7\% higher accuracy (68.26\% vs. 67.50\%), with $11.5\times$ faster online inference time (0.628s vs. 7.20s). 

\begin{table}[!t]
    \centering
    \caption{CIFAR-100 Comparison}
    \resizebox{\columnwidth}{!}{
    \begin{threeparttable}
        \begin{tabular}{c | c c c c c}
            \toprule
            & \multirow{2}{*}{Methods} & \#ReLUs & Test Acc. & Online Lat. & \multirow{2}{*}{Acc./ReLU} \\ 
            & & (K) & (\%) & (s) & \\
            \midrule
            \multirow{11}{*}{\rotatebox[origin=c]{90}{\small ReLU $\leq$ 100K}} & SNL\textsuperscript{\#} & 12.9 & 66.53 & 0.291 & 5.517 \\
            & SNL\textsuperscript{\#} & 15.0 & 67.17 & 0.334 & 4.478 \\
            & SNL\textsuperscript{\#} & 24.9 & 70.05 & 0.542 & 2.813 \\
            & SNL\textsuperscript{\#} & 49.9 & 73.75 & 1.066 & 1.478 \\
            & DeepReDuce & 12.3 & 64.97 & 0.45 & 5.282 \\
            & DeepReDuce & 28.7 & 68.68 & 0.56 & 2.393 \\
            & DeepReDuce & 49.2 & 69.50 & 1.19 & 1.413 \\
            & Sphynx & 25.6 & 66.13 & 0.727 & 2.583 \\
            & Sphynx & 51.2 & 69.57 & 1.335 & 1.359 \\
            & Sphynx & 102.4 & 72.90 & 2.385 & 0.712 \\
            & CryptoNAS & 100.0 & 68.5 & 2.30 & 0.685 \\
            \midrule
            \multirow{7}{*}{\rotatebox[origin=c]{90}{\small ReLU $\leq$ 300K}} & SNL\textsuperscript{*} & 120.0 & 76.35 & 2.802 & 0.636 \\ 
            & SNL\textsuperscript{*} & 150.0 & 77.35 & 3.398 & 0.516\\
            & SNL\textsuperscript{*} & 180.0 & 77.65 & 4.054 & 0.431 \\
            & DeepReDuce & 197.0 & 75.51 & 3.94 & 0.383 \\
            & DeepReDuce & 229.4 & 76.22 & 4.61 & 0.332 \\
            & CryptoNAS & 344.0 & 76.0 & 7.50 & 0.221 \\
            & Sphynx & 230.0 & 74.93 & 5.12 & 0.326 \\
            \bottomrule 
        \end{tabular}
        \begin{tablenotes}
            \scriptsize
            \item \textsuperscript{\#} Starts with pretrained ResNet18.
            \item \textsuperscript{*} Starts with pretrained Wide-ResNet 22-8.
        \end{tablenotes}
    \end{threeparttable}}
    \label{tab:cifar100}
\end{table}

\paragraph{SNL outperforms structured pruning.} 
We also compare SNL to magnitude-based structured pruning. As discussed previously, structured pruning methods remove entire filters (and as a by-product also reduce ReLU counts). However, the loss of learnable parameters leads to significant losses in accuracy. We compare SNL, $\ell_1$-magnitude filter pruning~\cite{li2016pruning} and LFPS-filter pruning~\cite{He2019FilterPV} with pretrained Resnet-18 networks on CIFAR-10 and CIFAR-100. The results are available in \Figref{fig: pareto curve}. SNL outperforms $\ell_1$ pruning by a large margin on accuracy for similar ReLU counts. While LFPS-pruning is more accurate than $\ell_1$ pruning, SNL is able to achieve the same accuracy while only requiring $\sim4\times$ fewer ReLUs. 
See also Appendix~\ref{appsec:results} for several additional ablation studies.

\section{Theoretical analysis} 
\label{subsec: expressivity}

Figure~\ref{fig: C100 relu distribution} in Section~\ref{sec:exp} reveals interesting trends. The blue bars indicate the number of ReLUs in every layer of a (standard) ResNet-18 model. Since the network is contractive, this number generally decreases as a function of depth. But after applying \textsc{SNL}, the red bars indicate that across the entire range of ReLU target budgets, \emph{neurons in earlier layers are linearized at a far greater rate than later layers}. This aligns well with earlier findings; DeepReDuce~\cite{DEEPREDUCE} also hypothesized that (manually) culling all ReLUs in the first few layers provides the largest latency improvement with the smallest accuracy drop.

We now present theoretical evidence as to why this might be the case. One way to measure the effect of linearizing ReLUs in a network is to compare the \emph{expressivity} of networks before and after linearization. To measure the expressivity, we leverage the notion of \emph{memorization capacity}. 

\begin{definition}[\citet{yun2018small,vershynin2020memory}]
    Let $f_W$ be a neural network of a given architecture with learnable parameters $W$. The memorization capacity of $f_W$ is the largest $N$ such that the following condition is satisfied: For all inputs $\{x_i\}_{i=1}^N \subset \R^{d_x}$ and for all $\{y_i\}_{i=1}^N \subset [-1, 1]$, there exists a choice of parameters $W$ such that $f_{W}(x_i) = y_i$ for $1 \leq i \leq N$.
\end{definition}

Our main results focus on dense feed-forward neural networks (FNN) \footnote{Analogous extensions to convolutional networks with/without residual connections are interesting directions for future work.}. Under the assumption that the dataset $\{(x_i, y_i)\}_{i=1}^N$ contains distinct data points, \cite{yun2018small} provide an upper bound on memorization capacity of FNNs with 2 hidden layers.

\begin{theorem}[\citet{yun2018small}]
    Consider an FNN $f_W$ with two hidden layers and piecewise linear activation function $\sigma$. Suppose that the hidden layers have width $d_1$ and $d_2$ and that the number of pieces in the activation function is $p$. Then, the memorization capacity of $f_W$ is no greater than:
    $$ p(p-1)d_1d_2 + (p-1)d_2 + 2 \, . $$
    For the case of ReLU networks, $p=2$ and the above bound on the capacity simplifies to $2d_1 d_2 + d_2 + 2$.
    \label{theorem: memorization}
\end{theorem}
The intuition for this theorem is based on the fact that FNNs with ReLU activations implement piecewise affine functions; so an upper bound on their capacity can be obtained by counting the number of ``linear pieces" in the range of the FNN, which can be recursively computed for each layer. See the appendix of \citep{yun2018small} for a detailed proof.

Let us now imagine selectively linearizing such an FNN; more specifically, suppose we only linearize all but a fraction $\alpha_1$ of the neurons in the first layer, and all but a fraction $\alpha_2$ of the neurons in the second layer. Mirroring the same counting argument as above, we obtain the following corollary to Theorem~\ref{theorem: memorization}.

\begin{corollary}
    Consider an FNN $f_W$ with 2 hidden layers of width $d_1$ and $d_2$ respectively, and piecewise linear activation function $\sigma$ with $p$ pieces. Let $\alpha_i \in [0, 1], i \in \{1,2\}$ be the fraction of nonlinear activations retained at each layer, i.e., without loss of generality the $l$\textsuperscript{th} layer $k$\textsuperscript{th} neuron exhibits the activation:
    \begin{align*}
    a^l_k(x) = 
    \begin{cases}
        \sigma(z^l_k(x)) \quad \text{if } k \in [\alpha_l d_l] \\
        z^l_k(x) \quad  \text{if } k \in [d_l] \setminus [\alpha_l d_l]
    \end{cases}
    \end{align*}
    Then, the memorization capacity of $f_W$ is no greater than:
    \begin{align*}
        \alpha_1\alpha_2d_1d_2p(p-1) + \alpha_2d_2(p-1) + \\
        \alpha_1(1-\alpha_2) d_1d_2(p-1) + 2 \, .
    \end{align*}
    For ReLU networks, the above bound simplifies to: 
    $\alpha_1(1 + \alpha_2) d_1d_2 + \alpha_2d_2 + 2.$
    \label{corrollary: SNL memorization}
\end{corollary}
Corollary~\ref{corrollary: SNL memorization} shows that the memorization capacity of a selectively linearized network (obtained, say, by applying \textsc{SNL}) is strictly smaller than the original network whenever $\alpha_1$ or $\alpha_2$ is smaller than 1.

It is useful to see how selective linearization compares with {(structured) pruning}. Suppose that instead of just dropping the ReLUs activation, we eliminate the entire neuron from the network, i.e., the network widths now become $\alpha_1 d_1$ and $\alpha_2 d_2$ respectively. The following corollary can be obtained by directly instantiating these in Theorem~\ref{theorem: memorization}.

\begin{corollary}
    Consider a 3-layer FNN $f_W$ with piecewise linear activation $\sigma$ with $p$ pieces. Let $\alpha_1, \alpha_2 \in [0, 1]$ be the fraction of nonlinear activations retained at the first and second layers, respectively. Then, the memorization capacity of $f_W$ is no greater than:
    $$\alpha_1\alpha_2d_1d_2p(p-1) + \alpha_2d_2(p-1) + 2 < N .$$
    For ReLU networks, the above bound simplifies to $2 \alpha_1 \alpha_2 d_1 d_2 + \alpha_2 d_2 + 2.$
    \label{corollary: SP memorization}
\end{corollary}

Notice the gap between the capacity upper bounds in Corollary~\ref{corrollary: SNL memorization} and Corollary~\ref{corollary: SP memorization}. Specifically, selective linearization enjoys an additional additive term:
$$ \alpha_1 (1 - \alpha_2) d_1 d_2, $$
compared to pruning. In fact, equality between the capacity bounds is achieved when $\alpha_2 = 1$, i.e., when all the neurons in the second layer are retained. 

This observation might indicate that retaining most ReLUs (i.e., driving $\alpha_2 \rightarrow  1$ in the second layer is the key. In fact, we can go further. From Corollary~\ref{corrollary: SNL memorization}, if we fix an overall ReLU budget for the network, then we can derive optimal conditions on the fraction parameters $\alpha_1$ and $\alpha_2$ to maximize memorization capacity. To simplify notation, we restrict our attention to ReLU activations (i.e., $p=2$). 

\begin{theorem}
    Consider a 3-layer FNN with ReLU activation. Let $\alpha_1, \alpha_2 \in [0, 1]$ be the fraction of nonlinear activations retaiend at the first and second hidden layers, respectively. Given a ReLU budget $B > d_2$, the choice $\alpha_1=\frac{B+d_2-1}{2d_1}$ and $\alpha_2=\frac{B-d_2+1}{2d_2}$ maximizes the memorization capacity. 
    \label{corollary: optimal relu distribution}
\end{theorem}

This result confirms that the optimal ``retention ratio'' for each network is inversely proportional to its width, and therefore in contractive networks (such as those typical in classification) where $d_1 > d_2$, linearizing a greater number of neurons in the first layer is beneficial. As a concrete example, consider a 3-layer FNN with $d_1 = 50000$ and $d_2 = 5000$, and budget $B=10000$. Then, the optimal choices are $\alpha_1 = 14999/100000 \approx 0.15$ and $\alpha_2 = 5001/10000 \approx 0.5$. Therefore, linearizing $42,500$ neurons in the first layer, and only $7,500$ neurons with ReLU in the second layer gives the best memorization capacity among all possible choices.

\section{Discussion}

We have developed a simple method to selectively linearize deep neural networks with ReLU activations, and using this method, demonstrated state-of-the-art networks for the problem of private inference.

Numerous directions for further work remain. Our work focused on reducing online runtime of private inference; a more holistic approach would consider both pre-processing and online costs. Instead of selectively linearizing certain ReLUs, replacing with other activations may be beneficial. Finally, deriving tighter bounds on memorization capacity for general architectures appears doable.

\section*{Acknowledgements}
This work was supported in part by the National Science Foundation (under grants CCF-2005804 and 1801495), USDA/NIFA
(under grant 2021-67021-35329), the Applications Driving Architectures (ADA) Research Center, a JUMP Center co-sponsored by Semiconductor Research Consortium (SRC) and the Defense Advanced Research Projects Agency (DARPA).

\bibliography{example_paper}
\bibliographystyle{icml2021}

\clearpage
\appendix
\onecolumn
\section{Proofs}

\subsection{Proof of Corollary~\ref{corrollary: SNL memorization}}


\begin{proof}
    The proof is based on the idea of counting the number of pieces in the range of networks with piecewise linear activations. Consider any vector $u \in \R^{d_x}$, and define the following dataset: $x_i = iu, y_i=(-1)^i$, for all $i \in [N]$. From \cite{telgarsky2015representation, yun2018small} we have:
    
    \begin{lemma}
        If $g: \R \to \R$ and $h: \R \to \R$ are piecewise linear with $k$ and $l$ linear pieces, respectively, then $g+h$ is piecewise linear with at most $k+l-1$ pieces, and $g \circ h$ is piecewise linear with at most $kl$ pieces. 
    \end{lemma}
    
    Fix $u$ and consider the output of layer 1 for input $x=tu$, where $t$ is a scalar variable: $\bar{a}^1(t) := a^1(tu)$. For each $j \in [\alpha_1d_1]$, $\bar{a}_j^1(\cdot)$ has at most $p$ pieces. On the other hand, $j^\prime \in [d_1] \setminus [[\alpha_1d_1]]$, $\bar{a}_{j^\prime}^1(\cdot)$ is a linear function and has $1$ piece. The input to layer 2 is the weighted sum of $\bar{a}_j^1(\cdot)$'s and $\bar{a}_{j^\prime}^1(\cdot)$'s. Each $\bar{z}_k^2(t) = z_k^2(tu)$ has $\alpha_1d_1(p-1)+1$ pieces. Now $\alpha_2d_2$ neurons after the activation $\sigma$ have at most $\alpha_1d_1p(p-1)+p$ pieces and $(1-\alpha_2)d_2$ neurons have at most $\alpha_1d_1(p-1)+1$ pieces. Therefore, the maximum number of pieces from the weighted sum of the second hidden layer neurons is calculated as:
    
    \begin{align*}
        &\alpha_1d_1p(p-1)+p+\sum_{i=1}^{\alpha_2d_2-1}(\alpha_1d_1p(p-1)+p-1) +  \sum_{i=1}^{(1-\alpha_2)d_2}(\alpha_1d_1(p-1)+1-1) \\
        &= \alpha_1\alpha_2d_1d_2p(p-1) + \alpha_2d_2(p-1) + \alpha_1(1-\alpha_2) d_1d_2(p-1) + 1 \\
    \end{align*}
    This calculation tells that a 3-layer network has at most $\alpha_1\alpha_2d_1d_2p(p-1) + \alpha_2d_2(p-1) + \alpha_1(1-\alpha_2) d_1d_2(p-1) + 1$ pieces. If this number is strictly smaller than $N-1$, the network cannot never perfectly fit the given dataset. 
\end{proof}

\subsection{Proof of Corollary~\ref{corollary: SP memorization}}
\begin{proof}
    The proof is a direct consequence of Theorem~\ref{theorem: memorization} by replacing $d_1$ and $d_2$ with $\alpha_1d_1$ and $\alpha_2d_2$.
\end{proof}

\subsection{Proof of Theorem~\ref{corollary: optimal relu distribution}}
\begin{proof}
    By setting $p=2$, we consider the following optimization problem:
    \begin{align*}
        &\max_{\alpha_1, \alpha_2} \alpha_1\alpha_2d_1d_2 + \alpha_1d_1d_2 + \alpha_2d_2 + 1 \, , \\
        &\text{s.t. } \alpha_1d_1 + \alpha_2d_2 = B \, .
    \end{align*}
    
    Constructing the Lagrangian, 
    \begin{align*}
        L(\alpha_1, \alpha_2, \lambda) = \alpha_1\alpha_2d_1d_2 + \alpha_1d_1d_2 + \alpha_2d_2 + 1 - \lambda(\alpha_1d_1 + \alpha_2d_2 - B) .
    \end{align*}
    By checking the first order condition of the Lagrangian, we get 
    $$\alpha_1 = \frac{B+d_2-1}{2d_1}, \qquad \alpha_2 = \frac{B-d_2+1}{2d_2} .$$
    
    We examine the determinant of the Hessian $H$ which is defined as
    \begin{equation*}
        H = \begin{bmatrix}
            0 & d_1 & d_2 \\
            d_1 & 0 & d_1d_2 \\
            d_2 & d_1d_2 & 0 
        \end{bmatrix} .
    \end{equation*}
    One can verify that the determinant $\det H > 0$ since $d_1, d_2 > 0$ and $\det H$ has the sign $(-1)^2=(-1)^{1+1}$ so the critical point $(\frac{B+d_2-1}{2d_1}, \frac{B-d_2+1}{2d_2})$ from the first order condition is indeed a maximum.
\end{proof}



\clearpage
\section{Additional Results}
\label{appsec:results}

\begin{figure*}[!t]
    \centering
    \resizebox{\textwidth}{!}{
    \begin{tabular}{c c c c}
         \pgfplotsset{compat=1.14}

\begin{tikzpicture}
\pgfplotsset{set layers}
\begin{axis}[
scale only axis,
xmin=0,xmax=377,
axis y line*=left,
ymin=30, ymax=80,
xlabel=Epochs,
ylabel style = {align=center},
ylabel={Test Acc. \ref{alpha 1e-5 test acc}},
grid=major,
grid style={dashed},
width=5cm
]
\addplot[mark=x,red, thick]
    table {tikz/alpha_1e-5_test_acc.csv}; 
\label{alpha 1e-5 test acc}
\end{axis}
\begin{axis}[
scale only axis,
xmin=0,xmax=377,
axis y line*=right,
axis x line=none,
ymin=0, ymax=500000,
ylabel style = {align=center},
ylabel={ReLU Count \ref{alpha 1e-5 test relu}},
width=5cm
]
\addplot[mark=*,blue, thick]
    table {tikz/alpha_1e-5_test_relu.csv}; 
\label{alpha 1e-5 test relu}

\end{axis}
\end{tikzpicture} & \pgfplotsset{compat=1.14}

\begin{tikzpicture}
\pgfplotsset{set layers}
\begin{axis}[
scale only axis,
xmin=0,xmax=57,
axis y line*=left,
ymin=30, ymax=80,
xlabel=Epochs,
ylabel style = {align=center},
ylabel={Test Acc. \ref{alpha 1e-4 test acc}},
grid=major,
grid style={dashed},
width=5cm
]
\addplot[mark=x,red, ultra thick]
    table {tikz/alpha_1e-4_test_acc.csv}; 
\label{alpha 1e-4 test acc}
\end{axis}
\begin{axis}[
scale only axis,
xmin=0,xmax=57,
axis y line*=right,
axis x line=none,
ymin=0, ymax=500000,
ylabel style = {align=center},
ylabel={ReLU Count \ref{alpha 1e-4 test relu}},
width=5cm
]
\addplot[mark=*,blue, ultra thick]
    table {tikz/alpha_1e-4_test_relu.csv}; 
\label{alpha 1e-4 test relu}

\end{axis}
\end{tikzpicture} & \pgfplotsset{compat=1.14}

\begin{tikzpicture}
\pgfplotsset{set layers}
\begin{axis}[
scale only axis,
xmin=0,xmax=11,
axis y line*=left,
ymin=30, ymax=80,
xlabel=Epochs,
ylabel style = {align=center},
ylabel={Test Acc. \ref{alpha 5e-4 test acc}},
grid=major,
grid style={dashed},
width=5cm
]
\addplot[mark=x,red, ultra thick]
    table {tikz/alpha_5e-4_test_acc.csv}; 
\label{alpha 5e-4 test acc}
\end{axis}
\begin{axis}[
scale only axis,
xmin=0,xmax=11,
axis y line*=right,
axis x line=none,
ymin=0, ymax=500000,
ylabel style = {align=center},
ylabel={ReLU Count \ref{alpha 5e-4 test relu}},
width=5cm
]
\addplot[mark=*,blue, ultra thick]
    table {tikz/alpha_5e-4_test_relu.csv}; 
\label{alpha 5e-4 test relu}

\end{axis}
\end{tikzpicture} & \pgfplotsset{compat=1.14}

\begin{tikzpicture}
\pgfplotsset{set layers}
\begin{axis}[
scale only axis,
xmin=0,xmax=10,
axis y line*=left,
ymin=30, ymax=80,
xlabel=Epochs,
ylabel style = {align=center},
ylabel={Test Acc. \ref{alpha 1e-3 test acc}},
grid=major,
grid style={dashed},
width=5cm
]
\addplot[mark=x,red, ultra thick]
    table {tikz/alpha_1e-3_test_acc.csv}; 
\label{alpha 1e-3 test acc}
\end{axis}
\begin{axis}[
scale only axis,
xmin=0,xmax=10,
axis y line*=right,
axis x line=none,
ymin=0, ymax=500000,
ylabel style = {align=center},
ylabel={ReLU Count \ref{alpha 1e-3 test relu}},
width=5cm
]
\addplot[mark=*,blue, ultra thick]
    table {tikz/alpha_1e-3_test_relu.csv}; 
\label{alpha 1e-3 test relu}

\end{axis}
\end{tikzpicture} \\
         (a) $\lambda=10^{-5}$ & (b) $\lambda=10^{-4}$ & (c) $\lambda=5\cdot10^{-4}$ & (d) $\lambda=10^{-3}$ \\ 
         \pgfplotsset{compat=1.14}

\begin{tikzpicture}
\pgfplotsset{set layers}
\begin{axis}[
scale only axis,
xmin=0,xmax=111,
axis y line*=left,
ymin=0, ymax=80,
xlabel=Epochs,
ylabel style = {align=center},
ylabel={Test Acc. \ref{lr 1e-4 test acc}},
grid=major,
grid style={dashed},
width=5cm
]
\addplot[mark=x,red, ultra thick]
    table {tikz/lr_1e-4_test_acc.csv}; 
\label{lr 1e-4 test acc}
\end{axis}
\begin{axis}[
scale only axis,
xmin=0,xmax=111,
axis y line*=right,
axis x line=none,
ymin=0, ymax=500000,
ylabel style = {align=center},
ylabel={ReLU Count \ref{lr 1e-4 test relu}},
width=5cm
]
\addplot[mark=*,blue, ultra thick]
    table {tikz/lr_1e-4_test_relu.csv}; 
\label{lr 1e-4 test relu}

\end{axis}
\end{tikzpicture} & \pgfplotsset{compat=1.14}

\begin{tikzpicture}
\pgfplotsset{set layers}
\begin{axis}[
scale only axis,
xmin=0,xmax=164,
axis y line*=left,
ymin=0, ymax=80,
xlabel=Epochs,
ylabel style = {align=center},
ylabel={Test Acc. \ref{lr 1e-3 test acc}},
grid=major,
grid style={dashed},
width=5cm
]
\addplot[mark=x,red, ultra thick]
    table {tikz/lr_1e-3_test_acc.csv}; 
\label{lr 1e-3 test acc}
\end{axis}
\begin{axis}[
scale only axis,
xmin=0,xmax=164,
axis y line*=right,
axis x line=none,
ymin=0, ymax=500000,
ylabel style = {align=center},
ylabel={ReLU Count \ref{lr 1e-3 test relu}},
width=5cm
]
\addplot[mark=*,blue, ultra thick]
    table {tikz/lr_1e-3_test_relu.csv}; 
\label{lr 1e-3 test relu}

\end{axis}
\end{tikzpicture} & \pgfplotsset{compat=1.14}

\begin{tikzpicture}
\pgfplotsset{set layers}
\begin{axis}[
scale only axis,
xmin=0,xmax=24,
axis y line*=left,
ymin=0, ymax=80,
xlabel=Epochs,
ylabel style = {align=center},
ylabel={Test Acc. \ref{lr 1e-2 test acc}},
grid=major,
grid style={dashed},
width=5cm
]
\addplot[mark=x,red, ultra thick]
    table {tikz/lr_1e-2_test_acc.csv}; 
\label{lr 1e-2 test acc}
\end{axis}
\begin{axis}[
scale only axis,
xmin=0,xmax=24,
axis y line*=right,
axis x line=none,
ymin=0, ymax=500000,
ylabel style = {align=center},
ylabel={ReLU Count \ref{lr 1e-2 test relu}},
width=5cm
]
\addplot[mark=*,blue, ultra thick]
    table {tikz/lr_1e-2_test_relu.csv}; 
\label{lr 1e-2 test relu}

\end{axis}
\end{tikzpicture} & \pgfplotsset{compat=1.14}

\begin{tikzpicture}
\pgfplotsset{set layers}
\begin{axis}[
scale only axis,
xmin=0,xmax=100,
axis y line*=left,
ymin=0, ymax=80,
xlabel=Epochs,
ylabel style = {align=center},
ylabel={Test Acc. \ref{lr 1e-1 test acc}},
grid=major,
grid style={dashed},
width=5cm
]
\addplot[mark=x,red, ultra thick]
    table {tikz/lr_1e-1_test_acc.csv}; 
\label{lr 1e-1 test acc}
\end{axis}
\begin{axis}[
scale only axis,
xmin=0,xmax=100,
axis y line*=right,
axis x line=none,
ymin=0, ymax=500000,
ylabel style = {align=center},
ylabel={ReLU Count \ref{lr 1e-1 test relu}},
width=5cm
]
\addplot[mark=*,blue, ultra thick]
    table {tikz/lr_1e-1_test_relu.csv}; 
\label{lr 1e-1 test relu}

\end{axis}
\end{tikzpicture} \\
         (e) $lr=10^{-4}$ & (f) $lr=10^{-3}$ & (g) $lr=10^{-2}$ & (h) $lr=10^{-1}$
    \end{tabular}}
    \caption{Ablation studies on Lasso coefficient $\lambda$ and learning rate $lr$. We test on ResNet18 architecture with CIFAR-100. (a)-(d) show the differences in behavior on test accuracy and ReLU counts with various Lasso coefficient values given the learning rate $lr=0.001$ and ADAM optimizer. High initial Lasso coefficients (as shown in (c) and (d)) induces the sparsity on $C$ rapidly and consequently the architecture's performance rapidly degrades as well. Small initial lasso coefficients from (a) and (b) gradually induces the sparsity on $C$ allowing the $W$ to adapt as well in order to maintain the networks' performance. (e)-(f) tracks the test accuracy and ReLU count given the Lasso coefficient $10^{-5}$. We observe the similar trend in learning rate compared to Lasso coefficients; using a small learning rate ((e), (f)) helps to gradually reduce ReLU counts and maintain the test accuracy.}
    \label{fig: ablation studies}
\end{figure*}
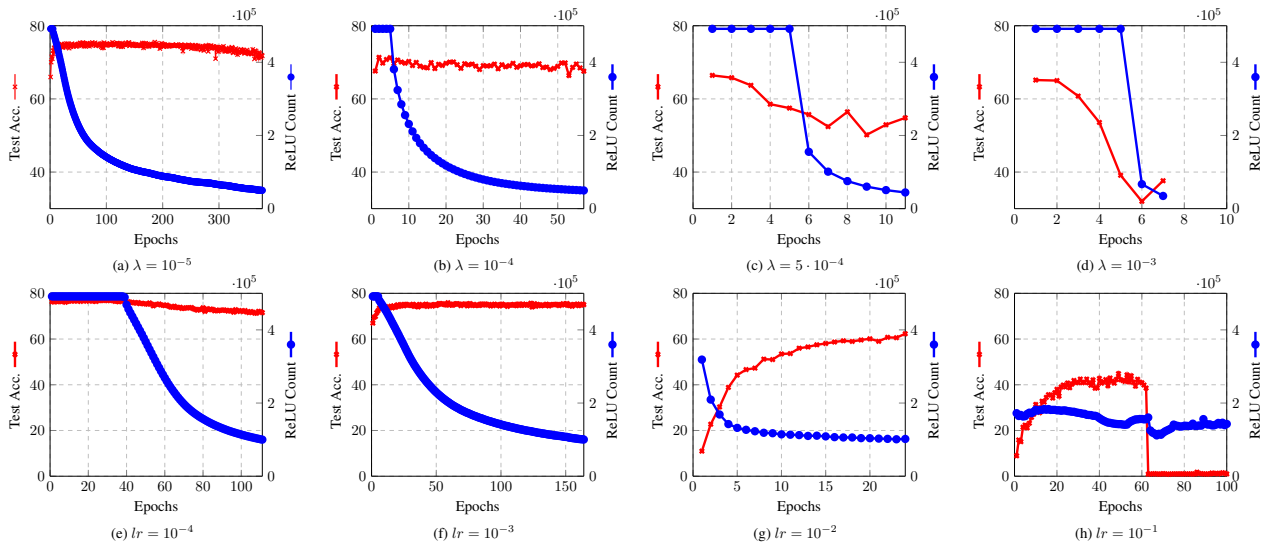

\subsection{Ablation Studies on Hyperparameters} We study the ablation studies on the primary hyperparameters: sparsity factor $\alpha$ and learning rate $lr$. We choose the network ResNet18 with the CIFAR-100 dataset. We first examine the role of the sparsity factor $\alpha$. The targeted ReLU budget here is $B=50000$, and the plots show the change of ReLU counts and test accuracy along with the training (corresponds to Algorithm~\ref{alg:finetune} line 4-8). We use an ADAM optimizer with a learning rate of $0.001$.

Along with the simultaneous training on the model parameter $W$ and auxiliary parameter $C$, the ideal situation will gradually reduce ReLU counts while maintaining the test accuracy close to the baseline pretrained model. We observe that increasing $\alpha$ induces the sparsity on $C$ faster; however, the test accuracy rapidly drops. On the other hand, the small $\alpha$ (e.g., first and second plot) gradually reduces the ReLU count while almost preserving the test accuracy. We heuristically observed incrementing the Lasso coefficient (line 5-6 in Algorithm~\ref{alg:finetune}) allows reaching the desired sparsity level on $C$ with Lasso coefficients initialized with small values.

We also examine the role of learning rate. We use the equivalent hyperparameters as above but with different targeted ReLU budget $B=100000$ and initial sparsity factor $\alpha=0.00001$. We observed that high learning rate such as $0.1$ or $0.01$ loses the performance in the beginning of the training and struggles to recover the performance compare to the original baseline performance (around 76\%). Our empirical results show that SNL algorithm works well with lower learning rate such as $0.001$ and $0.0001$. SNL algorithm with $lr=0.001$ terminates the simultaneous training with the test accuracy (before the finetuning) $75.04\%$ while $lr=0.0001$ achieves the $71.55\%$.

\subsection{Ablation Studies on Algorithmic Components} 

We study the variant of SNL algorithm by changing the algorithmic components. First, we consider zeroing out non-activated output instead of passing through the pre-activated inputs. In this case, the output, $\sigma_{c^i}(\cdot)$ at the i\textsuperscript{th} layer, can be expressed as:
\begin{equation}
    a_{c^i}^i = c^i \odot \sigma(z^i) + (\mathbf{1} - c^i) \odot \mathbf{0}
    \label{eq: feature map sparsification}
\end{equation}
Equation~\ref{eq: feature map sparsification} is equivalent to feature map sparsification where feature maps are zeroed out pixel-wise rather than filter-wise. Second, we perform the SNL algorithm from randomly initialized network (remark: SNL starts from the pretrained network).

\begin{figure}[!t]
    \centering
    \begin{tikzpicture}
    \begin{semilogxaxis}[legend pos=south east,
        legend style={nodes={scale=0.62, transform shape}},
        width=\columnwidth,
        height=6.2cm,
        xlabel= \small ReLU Count (K),
        ylabel=\small Test Accuracy (\%),
        xlabel style={at={(0.5, 0.1)}},
        ylabel style={at={(-0.05, 0.5)}},
        xmin=10, xmax=110,
        ymin=55,ymax=80,
        ylabel near ticks,
        xlabel near ticks,
        axis background/.style={fill=blue!0},
        grid=both,
        log basis x = 2,
        /pgf/number format/1000 sep={\,},
        log ticks with fixed point,
        grid style={line width=.1pt, draw=gray!10},
        major grid style={line width=.2pt,draw=gray!50},
        ]
    \addplot[mark=*,mark size=2.5pt,blue, thick] 
        plot coordinates {
            (12.9, 66.53)
            (14.9, 67.17) 
            (25.0, 70.05)
            (29.8, 70.18)
            (40.0, 72.35)
            (49.4, 73.75)
            (60.0, 74.85)
            (69.8, 75.29)
            (79.1, 75.72)
            (99.9, 75.94)
        };
    \addlegendentry{\textbf{SNL} (original)}


    

    \addplot[mark=x,mark size=2pt,red, thick] 
        plot coordinates {
            (12.9, 40.73)
            (14.9, 44.32) 
            (25.0, 56.28)
            (29.8, 61.15)
            (40.0, 64.99)
            (49.4, 68.15)
            (60.0, 68.12)
            (69.8, 71.1)
            (79.1, 72.27)
            (99.9, 73.18)
        };
    \addlegendentry{SNL-Zero Out}


    \addplot[mark=pentagon, mark size=2pt, brown, thick] 
        plot coordinates {
            (12.9, 65.06)
            (14.9, 66.35) 
            (25.0, 69.23)
            (29.8, 70.99)
            (40.0, 71.31)
            (49.4, 71.82)
            (60.0, 72.67)
            (69.8, 72.53)
            (79.1, 73.19)
            (99.9, 73.08)
        };
    \addlegendentry{SNL-Scratch}
    
    \addplot[black, thick] 
        plot coordinates {
            (12.9, 76.95)
            (99.9, 76.95)
        };
    \addlegendentry{ResNet18 (490K ReLU)}
    \end{semilogxaxis}
\end{tikzpicture}
    \caption{\textbf{Comparison between SNL and its variants on CIFAR-100}. All experiments in this plot uses ResNet18.}
    \label{fig: ablation comparison}
\end{figure}
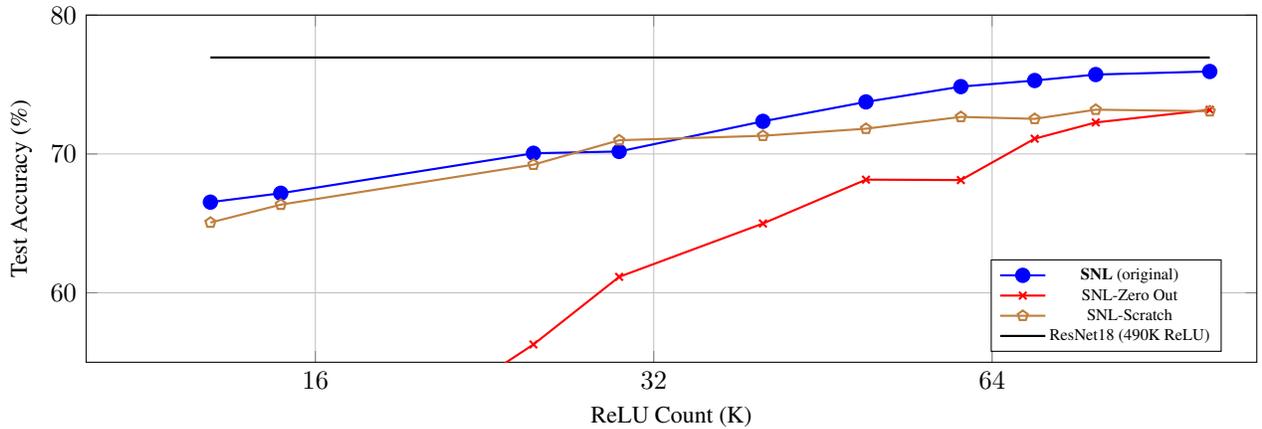

In Figure~\ref{fig: ablation comparison}, ``SNL-Zero-Out'' and ``SNL-Scratch'' correspond to feature map sparsification and the SNL network trained from scratch, respectively. For this comparison, we train a ResNet18 architecture with the CIFAR100 dataset. A black horizontal line is the reference performance ($76.95\%$ test accuracy) of the original ResNet18 network with $490K$ ReLU. We observe that the  ``SNL-Zero-Out'' performance degrades heavily in the sparse ReLU regime ($<50K$). Interestingly, ``SNL-Scratch'' showed a competitive performance to SNL while the original SNL consistently performed better in all cases except for one data point. Overall, this plot gives further evidence that the current SNL algorithm finds better performing architectures for a given ReLU budget over the SNL variants suggested by the reviewers. Finally, note that we incorporate knowledge distillation (KD) during the training, as clearly stated in the main paper (Line 268). We will add these results and discussion in the final draft.

\subsection{Role of the Lasso Coefficient}

\begin{figure}[!t]
    \centering
    \begin{tikzpicture}
%
\begin{axis}[legend pos=north east,
        legend style={nodes={scale=0.7, transform shape}},
        width=\columnwidth,
        height=6.0cm,
        xlabel=Epochs,
        ylabel style = {align=center},
        ylabel={ReLU Count ($\|\mathbf{c}\|_0$)},
        xmin=0, xmax=200,
        ymin=0,ymax=500000,
        ylabel near ticks,
        xlabel near ticks,
        axis background/.style={fill=blue!0},
        grid=both,
        grid style={line width=.1pt, draw=gray!10},
        major grid style={line width=.2pt,draw=gray!50},
        ]
\addplot[red, thick]
    table {tikz/alpha_1e-5_relucount.csv};
\addlegendentry{Lasso $\lambda=1\cdot10^{-5}$}

\addplot[yellow, thick]
    table {tikz/alpha_3e-5_relucount.csv};
\addlegendentry{Lasso $\lambda=3\cdot10^{-5}$}

\addplot[purple, thick]
    table {tikz/alpha_5e-5_relucount.csv};
\addlegendentry{Lasso $\lambda=5\cdot10^{-5}$}

\addplot[blue, thick]
    table {tikz/alpha_1e-4_relucount.csv};
\addlegendentry{Lasso $\lambda=1\cdot10^{-4}$}

\addplot[green, thick]
    table {tikz/alpha_5e-4_relucount.csv};
\addlegendentry{Lasso $\lambda=5\cdot10^{-4}$}

\addplot[orange, thick]
    table {tikz/alpha_1e-3_relucount.csv};
\addlegendentry{Lasso $\lambda=1\cdot10^{-3}$}

\end{axis}

\end{tikzpicture}
    \caption{\textbf{Effect of the Lasso Coefficient on supports of $\mathbf{c}$.} We track the number of supports in the auxiliary parameter $\mathbf{c}$ for various Lasso coefficients. See that the magnitude of Lasso coefficient ($\lambda$) determines the convergence rate and the support of $\mathbf{c}$.}
    \label{fig: ablation lasso convergence}
\end{figure}
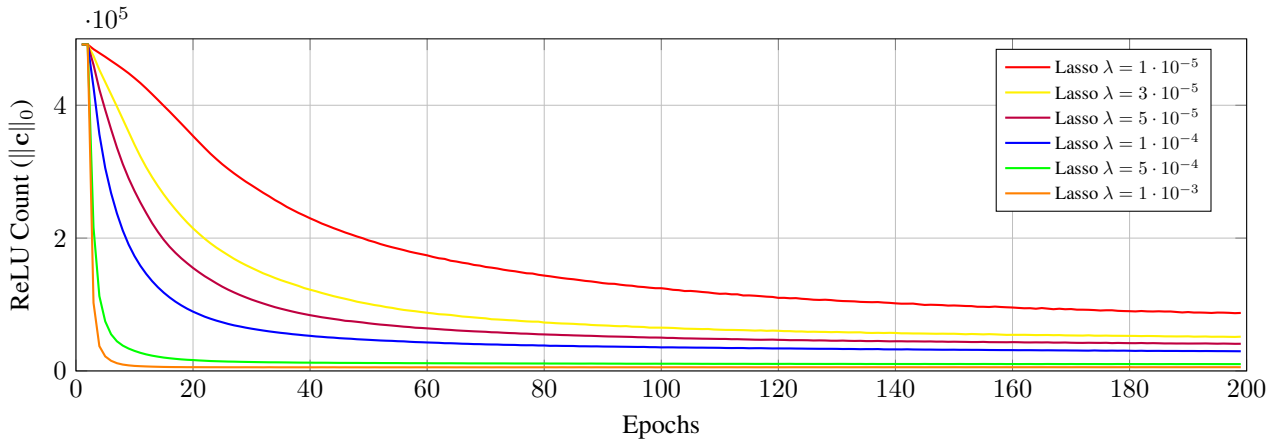

From our empirical observations, the magnitude of the Lasso coefficient ($\lambda$) determines the final (anti) sparsity of the slope parameters, and the convergence rate. We conduct experiments with a pre-trained ResNet18 architecture trained on CIFAR-100. We modify the SNL algorithm by removing the ``if-else'' condition, gradually increasing the Lasso coefficient, and track ReLU counts given by the auxiliary parameter $\mathbf{c}$ until $200$ epochs. The other hyperparameters remain the same as our reported experiments: ADAM, $lr=0.001$, weight decay of 0.0005. Figure~\ref{fig: ablation lasso convergence} shows that the magnitude of the Lasso coefficient directly contributes to the size of the support of $\mathbf{c}$ and the convergence rate. These observations motivate Line 6-7 in SNL, which uses homotopy-style optimization to gradually increases the Lasso coefficient to reach the desired sparsity while ensuring mild accuracy drop (also see Figure~\ref{fig: ablation studies} in the main paper showing that large $\lambda$ hurts test accuracy).

\subsection{Additional PI Results} Table~\ref{tab:cifar10} and Table~\ref{tab:tiny-imagenet} shows the PI comparison on CIFAR-10 and Tiny-ImageNet, respectively. Table~\ref{tab:resnet34} includes the additional PI results with pretrained ResNet34 network. Table~\ref{tab: structured SNL} contains the SNL results with channel-wise method demonstrated in Section~\ref{sec:exp}.

\begin{table}
    \centering
    \caption{CIFAR-10 Comparison.}
    \resizebox{0.6\columnwidth}{!}{
    \begin{threeparttable}
        \begin{tabular}{c | c c c c}
            \toprule
            & \multirow{2}{*}{Methods} & \#ReLUs & Test Acc. & {Acc./ReLU} \\ 
            & & (K) & (\%) & \\
            \midrule
            \multirow{6}{*}{\rotatebox[origin=c]{90}{\small ReLU $\leq$ 100K}} & SNL\textsuperscript{\#} & 12.9 & 88.23 & 6.840 \\
            & SNL\textsuperscript{\#} & 25.0 & 90.88 & 3.635 \\
            & SNL\textsuperscript{\#} & 60.0 & 92.63 & 1.544 \\
            & DeepReDuce & 36.0 & 88.5 & 2.458\\ 
            & DeepReDuce & 80.0 & 90.5 & 1.131\\
            & CryptoNAS & 86.0 & 91.28 & 1.061 \\
            \midrule
            \multirow{2}{*}{\rotatebox[origin=c]{90}{\small $\leq$ 400K}} & SNL\textsuperscript{*} & 240.0 & 94.24 & 0.4712 \\
            & SNL\textsuperscript{*} & 300.0 & 95.06 & 0.317 \\
            & CryptoNAS & 344.0 & 94.04 & 0.273 \\
            \bottomrule 
        \end{tabular}
        \begin{tablenotes}
            \scriptsize
            \item \textsuperscript{\#} Starts with pretrained ResNet18.
            \item \textsuperscript{*} Starts with pretrained Wide-ResNet 22-8.
        \end{tablenotes}
    \end{threeparttable}}
    \label{tab:cifar10}
\end{table}

\begin{table}
    \centering
    \caption{Tiny-ImageNet Comparison}
    \resizebox{0.6\columnwidth}{!}{
    \begin{threeparttable}
        \begin{tabular}{c | c c c c c}
            \toprule
            & \multirow{2}{*}{Methods} & \#ReLUs & Test Acc. & Online Lat. & \multirow{2}{*}{Acc./ReLU} \\ 
            & & (K) & (\%) & (s) & \\
            \midrule
            \multirow{5}{*}{\rotatebox[origin=c]{90}{\small ReLU $\leq$ 100K}} & SNL\textsuperscript{\#} & 59.1 & 54.24 & 1.265 & 0.918 \\
            & SNL\textsuperscript{\#} & 99.6 & 58.94 & 2.117 & 0.592 \\
            & DeepReDuce & 57.35 & 53.75 & 1.85 & 0.937 \\
            & DeepReDuce & 98.3 & 55.67 & 2.64 & 0.566 \\
            & Sphynx & 102.4 & 48.44 & 2.350 & 0.473 \\
            \midrule
            \multirow{5}{*}{\rotatebox[origin=c]{90}{\small ReLU $\leq$ 300K}} & SNL\textsuperscript{\#} & 198.1 & 63.39 & 4.183 & 0.320 \\
            & SNL\textsuperscript{\#} & 298.2 & 64.04 & 6.281 & 0.215 \\
            & DeepReDuce & 196.6 & 57.51 & 4.61 & 0.293 \\
            & DeepReDuce & 393.2 & 61.65 & 7.77 & 0.157 \\
            & Sphynx & 204.8 & 53.51 & 4.401 & 0.261 \\
            \midrule
            \multirow{3}{*}{\rotatebox[origin=c]{90}{\small $\leq$ 1000K}} & SNL\textsuperscript{*} & 488.8 & 64.42 & 10.281 & 0.132 \\
            & DeepReDuce & 917.5 & 64.66 & 17.16 & 0.070 \\
            & Sphynx & 614.4 & 60.76 & 12.548 & 0.099 \\
            \bottomrule 
        \end{tabular}
        \begin{tablenotes}
            \scriptsize
            \item \textsuperscript{\#} Starts with pretrained ResNet18.
            \item \textsuperscript{*} Starts with pretrained Wide-ResNet 22-8.
        \end{tablenotes}
    \end{threeparttable}}
    \label{tab:tiny-imagenet}
\end{table}

\begin{table}
    \centering
    \caption{SNL on ResNet34 Networks on CIFAR-100, Tiny-ImageNet}
    \resizebox{0.6\columnwidth}{!}{
    \begin{threeparttable}
        \begin{tabular}{c | c c c c c}
            \toprule
            & \multirow{2}{*}{Methods} & \#ReLUs & Test Acc. & Online Lat. & \multirow{2}{*}{Acc./ReLU} \\ 
            & & (K) & (\%) & (s) & \\
            \midrule
            \multirow{7}{*}{\rotatebox[origin=c]{90}{\small CIFAR-100}} & SNL & 14.9 & 67.08 & 0.339 & 4.502 \\
            & SNL & 25.0 & 69.68 & 0.551 & 2.787 \\
            & SNL & 30.0 & 70.99 & 0.656 & 2.366 \\
            & SNL & 50.0 & 72.91 & 1.075 & 1.458 \\
            & SNL & 80.0 & 74.19 & 1.705 & 0.927 \\
            & SNL & 99.9 & 74.76 & 2.122 & 0.748\\
            & SNL & 118.0 & 75.32 & 2.502 & 0.638  \\ 
            & SNL & 197.1 & 76.03 & 4.161 & 0.385\\
            \midrule
            \multirow{4}{*}{\rotatebox[origin=c]{90}{\scriptsize Tiny-ImageNet}} & SNL & 200.0 & 62.49 & 4.231 & 0.312 \\
            & SNL & 300.0 & 63.99 & 6.329 & 0.213 \\
            & SNL & 400.0 & 65.31 & 8.426 & 0.163 \\
            & SNL & 500.0 & 65.34 & 10.524 & 0.131 \\
            \bottomrule 
        \end{tabular}
    \end{threeparttable}}
    \label{tab:resnet34}
\end{table}

\begin{table}
    \centering
    \caption{CIFAR-100 Results on Channel-wise SNL.}
    \resizebox{0.6\columnwidth}{!}{
    \begin{threeparttable}
        \begin{tabular}{c c c c c}
            \toprule
            \multirow{2}{*}{Methods} & \#ReLUs & Test Acc. & Online Lat. & \multirow{2}{*}{Acc./ReLU} \\ 
            & (K) & (\%) & (s) & \\
            \midrule
            Channel-wise SNL\textsuperscript{\#} & 29.0 & 68.26 & 0.628 & 2.354 \\
            Channel-wise SNL\textsuperscript{\#} & 39.3 & 70.52 & 0.843 & 1.794 \\
            Channel-wise SNL\textsuperscript{\#} & 49.8 & 71.56 & 1.065 & 1.437 \\
            Channel-wise SNL\textsuperscript{*} & 77.7 & 73.47 & 1.650 & 0.940 \\
            Channel-wise SNL\textsuperscript{*} & 117.3 & 74.33 & 2.481 & 0.634 \\
            Channel-wise SNL\textsuperscript{*} & 200.0 & 77.45 & 4.216 & 0.387 \\
            \bottomrule 
        \end{tabular}
        \begin{tablenotes}
            \scriptsize
            \item \textsuperscript{\#} Starts with pretrained ResNet18.
            \item \textsuperscript{*} Starts with pretrained Wide-ResNet 22-8.
        \end{tablenotes}
    \end{threeparttable}}
    \label{tab: structured SNL}
\end{table}

\end{document}